# Optical Wake-up from Power-off State for Autonomous Optical Sensor Nodes

Uliana Dudko, Ludger Overmeyer

**Abstract**—Wireless sensor nodes spend most of their time in standby mode and wake up periodically to send the measurement data. Conventionally, the wake-up function is realized using a radio-frequency oscillator with an amplifier to recognize an activating radio signal. Therefore, even during standby operation, the sensor node utilizes a certain amount of energy, which can be critical for an energy-harvesting source. In this study, we propose a novel approach of an optical wake-up for autonomous sensor nodes, which employs a solar cell as a wake-up signal detector. The bright light flash coming from another node or a smartphone exposes a solar cell, which activates the sensor node. Unlike photodiodes or RF-antennas, solar cells do not require any additional energy to detect the light signal. Therefore, the proposed electric circuit allows the sensor node to wake-up from a complete power-off state. The solar cell of the novel wake-up receiver has a sensitive area of 8 mm x 10 mm. The wake-up signal can be recognized from a maximal distance of 25 cm at ambient illumination of 0 – 1600 lx with a transmitter optical power of 20 mW. At power-off state the power consumptions are the lowest among all existing off-the-shelf wake-up receivers: 248 pW at 0 lx and 627 nW at 1600 lx.

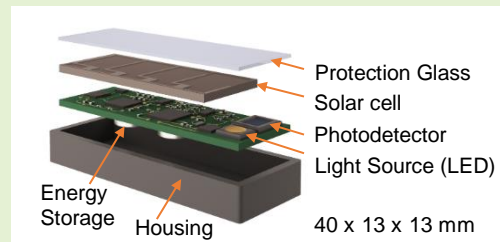

## I. Introduction

POWERING electronics from energy harvesting sources is a flexible and environmentally friendly solution, which provides long life-time operation and autonomy to various low-power devices, e.g. wireless sensors. They usually spend almost all their time in standby mode and are activated periodically by an external or internal wake-up signal. Commonly the external wake-up function is implemented using a radio frequency (RF) oscillator with an amplifier to recognize an activating radio signal. In this case, the always-on wake-up receiver in combination with a microcontroller constantly consumes the valuable energy harvested from ambient. The technologies, which can reduce the power consumption in standby mode, are of high importance. This study investigates the designs of an optical wake-up receiver, which is able to activate autonomous devices from the complete power-off state. Thus, no power for keeping a microcontroller of a wireless sensor in standby mode is required.

This research focuses on the implementation of the optical wake-up receiver for autonomous optical sensor nodes. These nodes communicate with each other using visible light and together establish a wireless sensor network. Each node employs a photodiode as a receiver and an LED as a transmitter. Power autonomy is achieved using a solar cell for optical energy harvesting. The main application of such nodes is data transfer in areas where RF communication is undesired. For instance, these networks can provide condition monitoring in hospitals or in specific production lines. Therefore, the reliable operation of the sensor nodes under indoor illumination conditions is required. The conceptual design of an autonomous optical sensor node is demonstrated in the abstract section. The special aspects of the communication between nodes and optical energy harvesting are discussed in our previous research [1, 2].

In this study, we examine the electronic designs for the sensor node activation from the power-off state. They are based on the optical wake-up triggered by a bright flash coming from another node or a smartphone. In the bright flash no key bit sequence is encoded. When the node is activated it communicates over a VLC link (LED - Photodiode) and performs the identification by a microcontroller. Therefore, the complexity and the power consumption of the wake-up circuit itself can be reduced. In this research we introduce two designs of wake-up receivers: one uses a solar cell and another one employs a phototransistor for optical wake-up detection. The main focus of this study is to implement the circuits using only off-the-shelf electronic components. Thus, the circuits are inexpensive to manufacture and are easy to reproduce.

This paragraph of the first footnote will contain the date on which the paper was submitted for review. This work was supported by the funding of Lower Saxony Ministry for Science and Culture (Germany) within the framework of the "Tailored light" project.

The authors are with the Institute for Transport and Automation Technology, Leibniz Universität Hannover, An der Universität 2, 30823 Garbsen, Germany (e-mail: uliana.dudko@ita.uni-hannover.de; ludger.overmeyer@ita.uni-hannover.de).

This article has supplementary downloadable material available at http://ieeexplore.ieee.org, provided by the authors. Digital Object Identifier: XX.XXXX/JSEN.2020.XXXXXXX

Section II explores the related work and advantages of the proposed wake-up technology in comparison with the prior work. In Section III the existing photodetectors and possible methods for optical wake-up from the power-off state are discussed. Section IV presents the circuit designs. Section V provides the evaluation and analysis of the results.

## II. RELATED WORK

Typically commercially available wake-up receivers based on RF technology consume standby power in the range of several microwatts [3–5]. Recently the first radio wake-up receiver with a total power draw of 7.6 nW was introduced in [6]. In the existing literature there are also several solutions based on acoustic signal recognition. Soliman et. al [7] presented an approach, where a piezoelectric acoustic resonator detects RF signals at the power consumption of 18 nW. In [8] an ultrasonic circuit was developed, which receives wireless wake-up acoustic signatures and consumes only 8 nW of power.

Scientists from the University of Michigan already proved in 2012 that the optical activation of sensor nodes requires essentially less energy compared to RF and acoustic approaches. In [9, 10] the researchers managed to reduce the power consumption of wake-up circuits to 695 pW and 380 pW respectively. The wake-up pulse detection in [9] is realized using a photovoltaic (PV) cell and does not have a dynamical adaption to the changes in ambient illumination. The approach of [10] is based on optical communication, where a photodiode in combination with a low-power amplifier circuit was employed as a wake-up detector. Although all mentioned studies present wake-up receivers operating in nanowatt and picowatt range, they are use a complex nanometer scale CMOS technology, which requires special expensive tools and clean room conditions for manufacturing and integration in communication systems.

Among off-the-shelf approaches there are examples of optical wake-up systems based on different photodetectors. In [11] the sensor nodes for vehicle detection in parking stations with a current drain of 5.5 µA were realized using LDRs. Mathews et. al [12] developed a free space optical receiver, where photodiodes identify laser wake-up signals from three directions. The proposed design consumes 83 µW. This number is reasonable since an amplification circuit built for a photodiode with separate commercial electronic components constantly requires tens of microwatts of energy. In [13] the researchers applied a solar cell for both energy harvesting and a wake-up. Although the idea of employing the solar cell is beneficial compared to the photodiode, since it does not require an additional amplifier, the circuit still consumes 95 µW. The reason for that is the integrated radio wake-up receiver, which takes a signal from the solar cell to the antenna input and constantly consumes several microwatts of power.

In this paper, we examine new approaches based on optical wake-up signal detection. We aim to reduce power consumption in standby mode to near-zero range by activating the node from the power-off state. This is the first study, which achieves this goal using a small-sized solar cell (8 mm x 10 mm) in conjunction with low-cost off-the-shelf electronic components.

## III. PHOTODETECTORS AND METHODS

There are four types of basic photodetectors, in which the electric properties depend on the amount of light incident on the active surface. These photodetectors are photodiodes, phototransistors, photoresistors, and PV cells. In this section we consider their properties and appropriateness to be employed as optical wake-up detectors in passive circuits.

A photodiode (PD) is a high impedance photosensitive device whose current amplitude is a linear function of received optical power. Since a PD can generate only a few millivolts, it is always necessary to use a transimpedance amplifier (TIA) to convert a photocurrent to a proportional output voltage [14]. The main component of TIA is an operational amplifier, which is an active device. The off-the-shelf ultra-low-power amplifiers constantly require at least several hundreds of nanoamps of supply current [15]. For this reason, it is not beneficial to employ a photodiode or PD arrays (e.g. image sensors) as wake-up detectors.

A phototransistor (PT) is a form of a bipolar transistor, which is virtually a photodiode with added base-emitter junction. The current in this junction is amplified to provide a much larger collector current that is hundreds of times greater than a PD [16]. In a circuit PT is commonly used in the switch mode, where it is either in "off" (cut-off) or "on" (saturated) state in response to the incident light. As a type of a bipolar transistor, PT has leakage or dark current which is typically in the nanoampere range [17]. In general, phototransistor switches have a relatively slow speed of response (from 5 to 90 µs depending on configuration [18]), which makes their use questionable in terms of short wake-up pulse detection.

A photoresistor or light-dependent resistor (LDR) is a passive light-sensitive device, which declines in resistance with respect to receiving luminosity on its surface. LDRs have time latency of changing its resistance in response to light fluctuations (approx. 10 ms to drop and 1.5 s to rise back [19]). Therefore, they are commonly used in an electronic circuit, where the instantaneous reaction to light is not required. In a bright environment the recognition of deviations in light intensity is difficult since there is only a slight drop in resistance. The resistance of LDRs at common illumination conditions (e.g. 300-500 lx) lies in the range of kilohms, which in certain circuit designs can cause current leakage of several microamperes.

A solar cell or PV cell is a device, which is optimized to convert light energy into electrical energy. In terms of working principle solar cells are similar to photodiodes but have a larger sensitive area. For this reason, solar cells have larger junction capacitance, which leads to relatively low time response (e.g. 1 ms for an area of 25 mm x 15 mm). Although the wake-up pulse must be in the range of milliseconds to be detected, the solar cell does not require an amplification circuit, which saves several microwatts of energy. Moreover, a solar cell can be theoretically used for both: wake-up detection and energy harvesting.

Generally, in this study we focus on the approaches of utilizing a solar cell for optical wake-up signal detection. Additionally, we examine one circuit with a phototransistor and a photoresistor for comparison with the solar cell design. The general idea of activating an autonomous optical sensor module

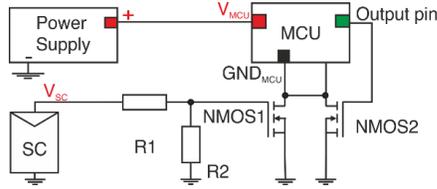

Fig. 2. The circuit diagram for the optical wake-up from power-off state using a solar cell.

from the power-off state using a solar cell as a wake-up detector is presented in Figure 2.

The bright light flash coming from another node or a smartphone exposes the solar cell. The voltage on the solar cell rises and switches the first N-channel MOSFET (NMOS1). A potential divider R1-R2 is used to control the solar cell response to illumination. The switched-on NMOS1 allows the current flow to the microcontroller. As soon as the microcontroller is activated, it sends the signal through an output pin to NMOS2 to hold the "on" state and check the validity of the wake-up signal. After performed measurement and data transfer, the microcontroller turns the NMOS2 off and shuts down.

The implementation of the circuit shown in Fig.2 brought up two main problems:
1) The microcontroller does not shut down completely and consumes a few milliwatts of power. The reason may be the internal indirect connection to the ground of some GPIOs of the microcontroller, for instance, through low-ohmic pull-down resistors.
2) The potential divider narrows the range of environmental illumination conditions in which the wake-up circuit responds appropriately to approx. 500 lx. This means that the wake-up receiver can function at room illumination of 400 lx, but it will no longer operate at 1000 lx (e.g. when the sun shines through a window). It is important to note that an exceeding room illumination above a certain level prohibits the functional capability. The solar cell gets saturated, the voltage raise on the solar cell is imperceptible and no wake-up can be detected. Therefore, the further wake-up designs are adjusted to the desired indoor light conditions of the typical range: from 0 lx to 1600 lx [20].

The solution for the first problem is to break the microcontroller from the positive supply using a P-channel MOSFET (PMOS) as shown in Fig.3. When NMOS1 is switched on, the voltage on PMOS1 gate is close to zero, so its $V_{gs}$ is close to $-V_{PMIC}$. The PMOS1 switches on and connects the microcontroller to the $V_{PMIC}$. The MOSFETs DMN65D8LFB and DMP21D5UFB4 are used as NMOS and PMOS respectively for all transistors in this and all the following designs due to the small dimensions (1 mm x 0.6 mm x 0.4 mm) and appropriate Vgs threshold characteristics. This circuit consumes 0.9 µA at 500 lx in standby mode.

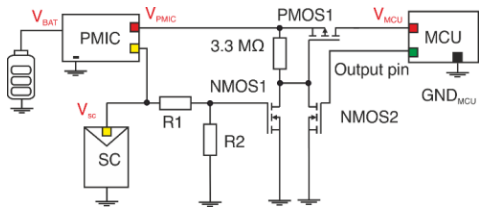

Fig. 3. Circuit schematics with the switch between the positive supply of the microcontroller $V_{MCU}$ and connection to the power management IC

In theory, one solar cell can be used for both wake-up detection and energy harvesting. Therefore, we connected its positive output to the power management IC (PMIC) BQ25570. The IC distributes the energy flow between the energy storage and the microcontroller. The boost converter with maximum power point tracking of the PMIC regulates the voltage on the output of the solar cell (Vsc) to maximize the charging efficiency. Therefore, during the charging process the PMIC fixes the Vsc to a certain voltage level. Thus, when a PMIC is connected, the solar cell cannot detect any of the transmitted wake-up pulses while the energy storage is charging (Fig.4).

In a simple case, when only a Schottky diode is used instead of a PMIC, the energy storage (e.g. a supercap) will drain the current from the solar cell until it is fully charged. For this reason, the voltage on the solar cell will be similar to the voltage on the energy storage Vbat. During the charging process the DC component of the Vsc (Schottky diode) constantly changes dependent on Vbat making it difficult to distinguish the peak-to-peak voltage of wake-up pulses. In order to detect the wake-up response in Vsc a complex amplification and signal processing circuit is required, which would essentially increase the power consumption of the wake-up receiver. The DC component of Vsc stabilizes when Vbat reaches the highest voltage on the solar cell or when the energy storage is fully charged. However, this state can easily change as soon as the room illumination increases, or the load discharges the storage for some time.

For this reason, we split the wake-up circuit from the energy-harvesting unit and use an extra solar cell as a wake-up signal detector.

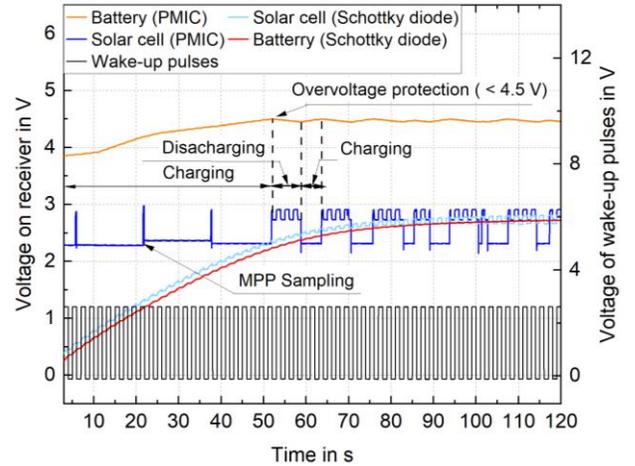

Fig. 4. Wake-up signal response on the solar cell for the circuit with energy harvesting using a PMIC and a Schottky diode

## IV. WAKE-UP CIRCUIT DESIGNS

The following designs we base on the circuit presented in Fig. 3 with the focus on improvement in ambient light immunity and signal integrity. The *first design* (Fig. 5) employs an amorthous solar cell AM-1456 (25 mm x 10 mm) for energy harvesting (SC2). One-third part of the same solar cell type is cropped using a laser cutter to form an area of 8 mm x 10 mm (SC1) for the wake-up detection. The average voltage generated on SC1 is 0.6 V and 0.75 V measured at 400 lx and 1600 lx respectively.

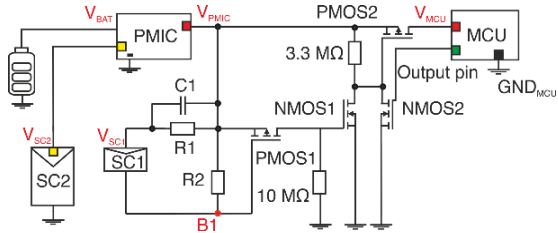

Fig. 5. Design 1: Extra solar cell (8 mm x 10 mm) as a wake-up detector

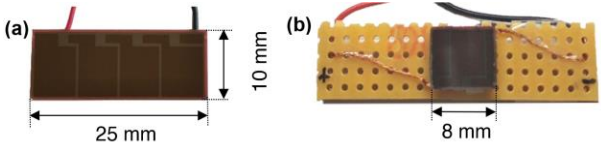

Fig. 6. (a) SC2 - solar cell for powering, (b) SC1 - solar cell for wake-up

The PMOS1 with high Vgs threshold (Vgs_th) is included to detect small deviations in voltage on SC1. The PMOS1-NMOS1 cascade serves the function of additional signal amplification. In general for PMOS, the Vgs and Vgs_th are negative values. If Vgs < Vgs_th, the transistor switches on, otherwise it is off. Thus, the higher Vgs_th is, the more sensitive is the PMOS. If the voltage raise on the SC1 is high and fast enough, the DC value that passes through the voltage divider, forms Vgs on the gate of PMOS1 lower than its Vgs_th. Thus, PMOS1 switches on, the voltage on gate of the NMOS1 rises to $V_{PMIC}$ and turns it on. The activated NMOS1 switches the PMOS2 on, which connects the microcontroller to the $V_{PMIC}$.

To the part with the potential divider we include a capacitor to provide the dynamical adaption to the changes in the room illumination. The capacitor C1 stores the constant voltage on the SC1 exposed to the ambient light. If the voltage on the solar cell rises slowly, C1 charges gradually. Thus, its potential increases the voltage on the gate of PMOS1 (B1), so the Vgs threshold of PMOS1 is not reached to turn it on. For this reason, the slow variations in the ambient illumination (e.g. the sun rises from behind a cloud) are not detected as wake-up signals. When a short light flash exposes the solar cell, the voltage on C1 cannot grow immediately, so the voltage on B1 falls, turns PMOS1 on, and connects the microcontroller to power.

The circuit's sensitivity to light influences two important criteria: the distance at which the wake-up signal can be received and the circuit power consumption. The more sensitive the circuit is, the higher is the activation distance and the higher is the current the circuit drains. Therefore, the combination of the resistances R1 and R2 of the potential divider, the capacitance value of the C1, and the threshold voltage of the PMOS1 must be chosen carefully. With the empirically selected values of R1 = 1.33 MΩ, R2 = 1.13 MΩ and C1 = 470 nF the best performance has been achieved.

The capacitor value adjusts the time $t_{amb}$ of how slow the ambient light should change to not be recognized as a wake-up signal. The time can be roughly estimated as $t_{amb} = 5\tau = R_E C$, where $\tau$ is a time constant of the capacitor, $C$ is the capacitance and $R_E$ is the equivalent resistance: $R_E^{-1} = (R_{sc}+R_1)^{-1} + R_2^{-1}$. Since the resistance of the solar cell $R_{sc}$ always slightly varies depending on the amount of light the solar cell was exposed to, $t_{amb}$ is not constant. In our circuit with the experimentally chosen R1, R2 and C1 we set $t_{amb}$ to be not less than 1 s for the illumination range defined in 3.2.

The 10 MOhm resistor between gate and source of the NMOS1 also influences the sensitivity of the circuit. It sets the current threshold, from which the NMOS1 must be switched on. The higher the value of this resistor, the less current over drain-source of PMOS1 is required to activate the NMOS1.

The *second design* employs a phototransistor PT1 for the optical wake-up signal detection as an alternative to the solar cell (Fig.7). In this circuit we use the phototransistor SFH3710 with the sensitive area of 0.29 mm$^2$. The role of the PT1 is to switch on when it is exposed to an activation flash. In order to provide the ambient light immunity, an LDR (GL5549) is used as a light sensor. By varying its resistance in response to the changes of ambient light the LDR adjusts the voltage on the gate of the NMOS1. Since the LDR has a time latency of changing its resistance, the slow variations in environmental illumination will not be detected as wake-up signals. At the same time, for a short light pulse, the LDR does not react immediately, which leads to a wake-up of the module.

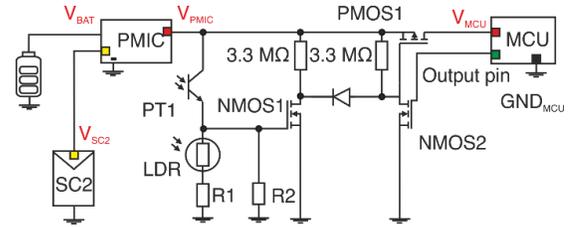

Fig. 6. Design 2: Phototransistor as a wake-up detector and LDR as a light sensor

The commercially available phototransistors and LDRs are very sensitive to any presence of light. Thus, even for ambient light levels of 500 lx, the PT1 would typically draw 100 μA. The LDR resistance drop at the examined light conditions (from 10 kΩ to 0.5 kΩ at 100-1600 lx) provides a very low rise in voltage, insufficient to detect a wake-up. In order to increase the resistance drop of the LDR (to hundreds of kΩ) and to reduce the sensitivity of the phototransistor, a black semitransparent coating foil is applied (Fig. 8).

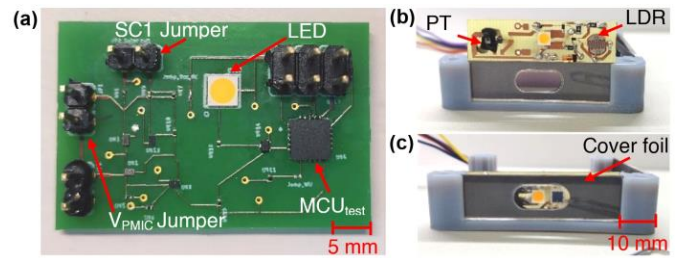

Fig. 7. The PCBs for testing the wake-up receivers (a) first design, second design without (b) and with cover foil (c)

In order to test the performance of both designs, the PCBs made of FR4 material are manufactured and assembled. Both PCBs are 1.6 mm thick with a 35 μm Cu layer. The first design prototype has extra stop lack cover for facilitating the soldering of small parts. The LED serves as indicator when the microcontroller is activated. The maximum distance between a transmitter and a wake-up detector at different light conditions for each design is further evaluated. Finally, the current consumption of both circuits is measured.

## V. Evaluation

In order to examine the maximum wake-up range of the designed circuits, we take measurements of the error rate at illumination levels from 0 to 1600 lx at a variable distance between an LED and a wake-up detector. The measurement setup (Fig. 8) includes an optocoupler to realize galvanic isolation between a measurement circuit and a wake-up receiver under test.

The evaluation process consists of the following steps.
1) The measuring microcontroller sends a wake-up pulse using an LED transmitter every 5 seconds.
2) The light signal captured by a detector (e.g. solar cell or phototransistor) activates the wake-up circuit, which then connects a microcontroller under test PIC16LF1509 (MCU$_{test}$) to 2.8 V.
3) As soon as MCU$_{test}$ is activated, it sends an indication signal to the optocoupler, which transfers the signal to the measuring microcontroller (MCU$_{meas}$).
4) MCU$_{meas}$ counts the amount of the transmitted and received wake-up signals and shows them on the LCD display.

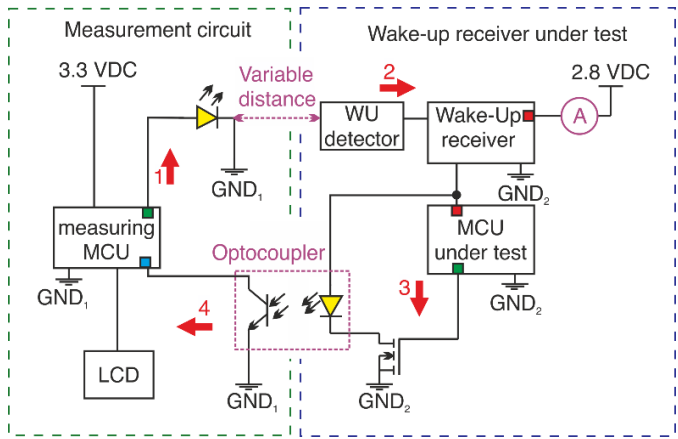

Fig. 8. Measurement setup

The LED transmitter employed for the measurements is the one that is integrated in the autonomous optical sensor module. It has a viewing angle φ of 120° and emits white light with an optical power of 20 mW. The LED is placed along the line-of-sight to the wake-up detector. The following diagrams show the number of errors out of 100 transmitted activating signals for three examined wake-up receiver designs.

The measurement results presented in Fig. 9 show that the first design provides a reliable wake-up scenario at illumination levels from 400 to 1600 lx at distances up to 25 cm. The detection is even possible at 2000 lx at small distances up to 5 cm. At 0 lx (e.g. at night) the highest distance of 30 cm has been achieved.

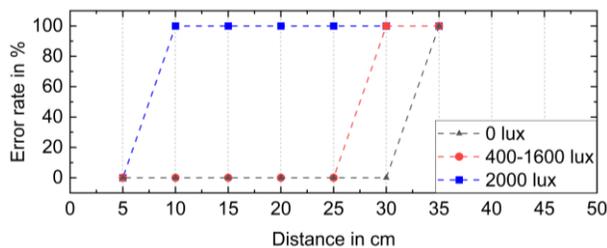

Fig. 9. Measurement of the maximal wake-up range for the first design. Wake-up detector: solar cell (8 mm x 10 mm)

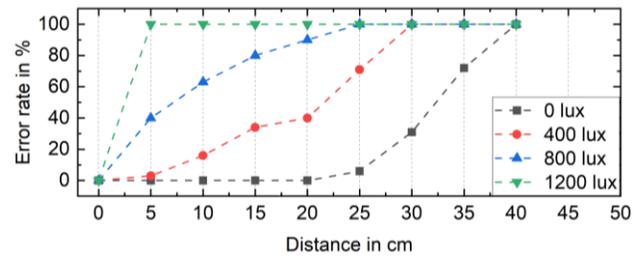

Fig. 10. Measurement of the maximal wake-up range for the second design. Wake-up detector: phototransistor

The detection distance is manly limited by the ambient light, which remains the major source of noise and at high illumination levels surpass the wake-up signal. Therefore, the reliable operation is only possible in a certain lighting range and can be adjusted in accordance to a particular application.

Another factors that limit the detection distance is the transmitter optical power and its viewing angle. In the conducted measurements an LED with broad radiation pattern φ=120° and, thus, high scattering is used (the worst case scenario). Depending on application the degree of directionality of the LED transmitter and the wake-up receiver can be improved using optical concentrators, which would essentially increase the communication distance.

The measurements shown in Fig. 10 reveal the inapplicability of the second design in terms of unreliable wake-up detection. Errors occur even at small distances between the LED and the phototransistor when ambient light is present. One reason for that behavior is the different response time of the phototransistor and the LDR at different ambient light conditions. When the LDR response is faster than the phototransistor response, the wake-up pulse cannot always be detected, as it happens at 800 lx (Fig.11.b).

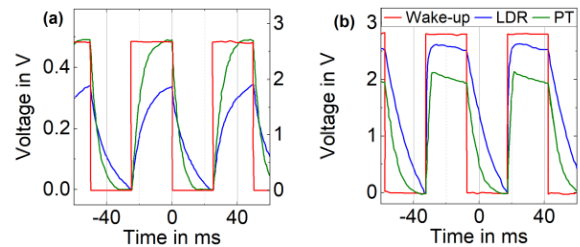

Fig. 11. Wake-up response of the phototransistor and LDR (a) at 400 lx, (b) at 800 lx

In order to examine the current consumptions of the designed wake-up receivers, the electrometer Keysight B2987A is connected as an ammeter in series to the 2.8 V power supply (Fig. 9). The measurements are conducted at a room temperature of 20° C. When measured at no illumination (0 lux) the wake-up receivers are placed in a fully-opaque enclosure. The measured current contains overall leakage of wake-up receiver circuit including the leakage of microcontroller. The results have shown that the first design consumes only 88.5 pA at 0 lx and 224 nA at 1600 lx (Fig. 12).

TABLE I
PERFORMANCE SUMMARY AND COMPARISON OF THE OFF-THE-SHELF WAKE-UP DESIGNS

| | [5] | [4] | [3] RF | [3] Optical | [11] | [12] | [13] | This work *first design* |
|---|---|---|---|---|---|---|---|---|
| Transmit method | RF | RF | RF | Blue and red LED | Shadowing | IR-LD | LED | White LED |
| WU detector | RF-antenna | RF-antenna | RF-antenna | PD | LDR | IR-PD | Solar cell | Solar cell |
| Standby power | 8.7 µW | 9.3 µW - 439.5 µW | 70.2 µW | ≥ 20.7 µA | 16.5 µW | 83 µW | 95 µW | 248 pW (0 lx)/ 627 nW (1600 lx) |
| Error free range | N/A | N/A | N/A | 4 m / 5 m @ 0 lx | ≥ 1 m | 15 m | 3 m / 16 m @ 0 lx | 0.25 m / 0.3 m @ 0 lx |
| RF sensitivity | -63.4 dBm | -77 dBm | -61 dBm | N/A | N/A | N/A | N/A | N/A |
| Transmitter power | N/A | N/A | N/A | 3 W | N/A | 0.7 mW @ φ = 6° | 87.9 mW | 20 mW @ φ = 120° |
| Total area | approx. 7.7 x 1.7 cm | 3.3 x 3.3 cm | N/A | N/A | N/A | approx. 8 × 7 cm | 3.6 × 2.6 cm plus PCB area | 0.8 × 1 cm plus 25 mm$^2$ of PCB area |
| WU signal | Bit code | Bit code | Bit code | Bit code | Light drop | Bit code | Bit code | Light pulse |

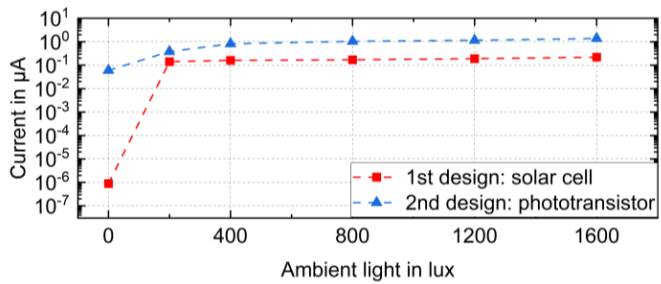

Fig. 12. Current consumption of designed wake-up receivers

The raise of current consumption with the illumination is due to the leakage through 10 MOhm and 3.3 MOhm resistors, when PMOS1 and NMOS1 are partially activated in response to the increasing voltage on the SC1.

The first wake-up receiver design was integrated into the autonomous optical sensor module. In Fig. 13 the two-layer PCB of such a module is depicted. All utilized resistors and capacitors have 0201 (0.6 x 0.3 mm) package size. As a result, the wake-up circuit excluding the solar cell occupies the area of only 5 mm x 5 mm.

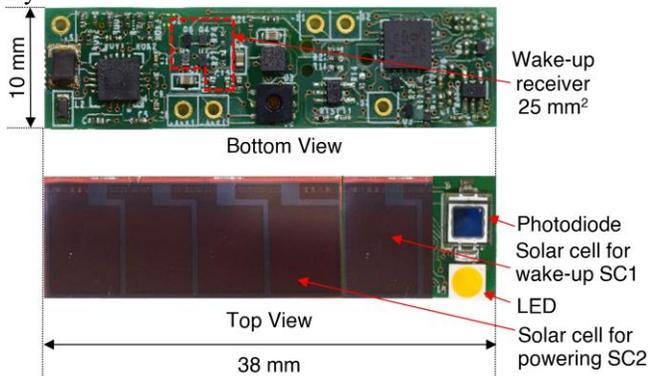

Fig. 13. Integrated wake-up receiver circuit in the main board of the autonomous optical sensor module

Table 1 compares the proposed optical wake-up receiver to prior work, showing the 13.8× and 10$^5$× improvement in standby power during the day light and at no illumination respectively. The implemented design occupies 1.05 cm$^2$ of two-layer PCB area and is the smallest among all existing off-the-shelf wake-up designs.

## VI. CONCLUSION

This study was carried out to design a power-efficient optical wake-up receiver, which activates an autonomous optical sensor module from the power-off state. The wake-up trigger is a bright flash, which comes from another sensor node or a smartphone. The key-recognition of a bit sequence is carried out by a microcontroller, when it is activated. Thus, the wake-up circuit can be simplified and the energy it consumes in standby mode can be minimized. The main focus of this research was to achieve the low power consumption and small dimensions of the wake-up receiver using exclusively off-the-shelf electronic components. Additionally, it was important to adapt the design for indoor illumination conditions in the range from 0 – 1600 lx. We introduced a general approach based on the solar cell as a wake-up detector and reveal the problem of using the same solar cell for energy harvesting. Two designs were presented, which were aimed to improve the general idea in terms of signal integrity and ambient light immunity. The wake-up detection of the first solution was based on the solar cells with sensitive areas of 8 mm x 10 mm. The second design employed a phototransistor and an LDR. The evaluation has shown that the first design has better performance compared to the second solution. Among the other off-the-shelf wake-up receives, the proposed circuit with a solar cell as detector has the lowest power consumption of 248 pW (0 lx) and 627 nW (1600 lx) and more compact area of 1.05 cm$^2$. The wake-up signal can be reliably obtained at distances of up to 25 cm at the transmitter optical power of 20 mW. This wake-up receiver can be further enhanced by integrating a key-recognition function. An integrated band-pass filter would allow the sensor node to be activated only when the pulse duration corresponds to a certain time interval.

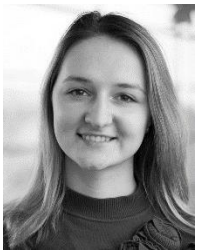
Uliana Dudko received the master's double degree in mechatronics from Leibniz University Hannover, Hannover, Germany, and Peter the Great St. Petersburg Polytechnic University, St. Petersburg, Russia, in 2016.

She is currently working towards the Ph.D. degree at the Institute for transport and Automation Technology, Leibniz University Hannover, since 2017. Her major research interests include low-power electronic design, internet of things, optical communication technologies, and optoelectronic packaging.

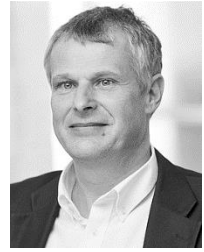
Ludger Overmeyer received the Dipl.-Ing. degree in electrical engineering from the Leibniz University Hannover, Hannover, Germany, in 1991.

He was a Research Associate and the Department Leader of Laser Zentrum, Hannover, from 1991 to 1997. He was the Leader for research and development with Mühlbauer AG, Roding, Germany. Since 2002, he has been a Professor and the Chair of the Institute of Transport and Automation Technology with the Leibniz University Hannover. He is currently the executive director of the Institute of Integrated Production GmbH (IPH) and also a member of the executive board of the Laser Zentrum Hannover. He has authored or coauthored over 100 conference papers, journal papers, and book chapters. His current research interests include transport technology, automation technology, laser materials processing, optoelectronic packaging, and planar optronic sensor systems.